# Particle size controlled magnetic loss in magnetite nanoparticles in RF-microwave region

Mudra Jadav, and S P Bhatnagar*

Department of Physics, Maharaja Krishnakumarsinhji Bhavnagar University, Bhavnagar-364001, India
*E-mail: spb@mkbhavuni.edu.in

**Abstract:** Frequency dependant complex magnetic permeability is used to understand RF-microwave behaviour of magnetic nanoparticles in the frequency range 250 MHz to 3 GHz. The stable dispersions of $Fe_3O_4$ nanoparticles with mean size varying between 11 to 16 nm are prepared for this purpose. The effect of mean particle size and external static magnetic field over microwave absorption properties of magnetic fluid is studied. It is observed that frequency of ferrimagnetic resonance ($f_{res}$), frequency of maximum absorption ($f_{max}$), loss tangent ($\tan \delta$) and reflection loss ($RL$) can be controlled by modifying mean particle size and strength of applied external static magnetic field. This kind of study can be useful for radio-microwave devices like tunable attenuator, EM sheilder, and other applications like Hyperthermia.

## I. INTRODUCTION

Magnetic fluids [1] are colloidal suspension of ferro/ferrimagnetic nanoparticles coated with a surfactant layer. Magnetic fluid is a smart material which responds to external magnetic field along with its fluid like properties. Magnetic fluids have large number of technological applications in various fields [2] [3] [4]. Some studies on frequency dependence of complex magnetic permeability and occurance of ferromagnetic resonance (FMR) are reported in [5-8] for magnetic fluids. Magnetic fluids are useful for designing radio-microwave devices due to their flexibility in shape and tunability with external magnetic field. Many of the workers highlighted their potential for radio-microwave applications such as in modulator [9], Electromagnetic shielding [10], controlled impedance device [11], insulator device [12], nonreciprocal device [13], hyperthermia [14] [15] & thermal recovery technique [16], and in microwave absorption & shielding using its composites [17][18]. Magnetic fluid parameters like particle size, shape, composition, surfactant, and non-magnetic carrier must be chosen to make them suitable for a particular application. The modificatios in these parameters can affect their properties. Modifications in particle size modify their properties like magnetic [19], rheological [20], optical [21] and microwave absorption properties [22] [23] of magnetic fluid. In the reports [22-23], researchers have measured ferromagnetic Resonance (FMR) in magnetic fluids at a fix frequency and applied magnetic field. They have obtained broader linewidth and lower resonance field for larger particles in comparision to the smaller ones. The effect of particle size over FMR and dispersion of resonance field was studied theoratically in [24][25]. In a recent report [26], researchers have studied temperature rise and specific absorption rate (SAR) at 126 kHz for $Fe_3O_4$ nanoclusters of varying size between 250 nm to 640 nm. They have shown that larger extent of temperature rise and SAR can be obtained for the nanocluster having highest saturation magnetization and largest crystallite size.

In our previous report [27], the effect of particle concentration, static magnetic field and it's orientation on complex magnetic permeability of magnetic fluid was studied. In this paper, the effect of particle size variation over complex magnetic permeability, microwave absorption, and reflection loss in magnetic fluid is reported. Broadband measurements in the frequency range 250 MHz to 3 GHz were carried out in contrast to the fixed frequency measurements reported earlier [22,23,26]. The microwave properties were studied as function of frequency as well as externally applied static magnetic field of strength 0 to 915 Oe. The fluid used was stable dispersion of single core $Fe_3O_4$ nanoparticles in contrast to the multi core $Fe_3O_4$ nanoclusters used in [26]. The field strengths used were comparable to anisotropy field ($H_A$) while in [22] [23], field strength used were much greater than $H_A$.

## II. EXPERIMENTAL

## A. Materials Preparation

The magnetite nanoparticles were synthesized by coprecipitation of two salt solutions $FeCl_3.6H_2O$ (SD fine chemicals) and $FeSO_4.7H_2O$ (SD fine chemicals) in the presence of 25% ammonia solution (Merck). Initially the mixture of two salt solutions was digested for 30 minutes at constant temperature and pH. The nanoparticles were coated with oleic acid (SD fine chemicals) surfactant and stabilized in low odor kerosene (SD fine chemicals) to prepare magnetic fluid. The magnetic fluid was centrifuged at 8000 RPM for 20 min in order to remove aggregates, if present. Different pH values were selected at a constant temperature for preparing magnetic fluids. These fluids were labelled as MF 1, MF 2, MF 3 and MF 4. Density of all fluids were 0.91 gm/cc. The complete method of preparation is discussed elsewhere [28].

## B. Methods

X-ray diffraction (Philips X'pert MPD System) was used for structural characterization of powder samples and diffraction data was analyzed by Reitveld refinement using the programme Materials Analysis Using Diffraction (MAUD) . The Transmission electron microscopy (TEM) (JEOL, JEM 2100) was used to determine particle size and size distribution. Open source software ImageJ was used for image analysis. The magnetization measurements were taken using search coil method. A search coil and compensating coil (2 cm long) were prepared by opposite winding of (36 SWG) wire with 500 turns on a nonmagnetic former with inner diameter of 1 cm. Both of these coils former were kept in an air core solenoid connected to power supply. The differential output from coils was measured by digital storage oscilloscope (Aplab D36025M). A glass tube containing known amount of magnetic fluid was quickly inserted into the search coil. The flux change was observed by peak signal on the oscilloscope screen. It's calibration was done using a magnetic fluid with known magnetization and calibration constant was obtained. The peak intensities for our sample fluids were converted into magnetization using calibration constant. The magnetization is detrmined as a function of magnetic field . Magnetic field was measured using digital gaussmeter with axial hall probe (SES Instruments Pvt. Ltd. DGM-204) Complex magnetic permeability of magnetic fluid was determined using Vector Network Analyzer (VNA) (Agilent 8714ES) in the frequency range 250 MHz to 3 GHz. The transmission/reflection technique [29] was used to measure the scattering parameters. Nicolson-Ross [30] and Weir [31] algorithm was used for calculation. A 50 Ω coaxial line cell was used as sample holder with 6.5 mm inner diameter and 15 mm outer diameter and 14 mm length. The coaxial line cell is made up of nonmagnetic material. VNA was calibrated and checked using known standards and known liquid. Measurements were taken under the static magnetic field with field strength between 0-915 Oe. The field was produced using an air core solenoid connected to a power supply and the sample holder was kept at centre of the solenoid's core. The schematic diagram of experimental set-up is shown in figure 1. The direction of static field was parallel to the cell axis and EM wave propagation direction. The blank measurement was done using air filled sample holder with and without external static magnetic field and it is confirmed that there is no effect of external magnetic field over the sample holder.

## III. Results and Discussion

Fig.2 shows the X-ray diffraction spectra for the nanoparticles refined using Reitveld refinement programme. The cubic crystal structure and spinel phase is confirmed by the X-ray spectra. For the inverse spinel arrangement of $Fe^{3+}$, $Fe^{2+}$ and $O^{2-}$ ions in magnetite, $O^{2-}$ ions occupy lattice sites, $Fe^{2+}$ ions occupy octahedral voids, half of the $Fe^{3+}$ ions occupy tetrahedral voids and the other half occupy octahedral voids. Electron spins of $Fe^{3+}$ ions at tetrahedral voids are aligned antiparallel to the electron spins of $Fe^{3+}$ ions at octahedral voids. The total magnetic moment from $Fe^{3+}$ ions is zero. Electron spins of $Fe^{2+}$ ions are aligned parallel to the spins of $Fe^{3+}$ ions at neighbouring octahedral voids. These are responsible for the net magnetization and ferrimagnetic nature [32] of magnetite. The crystallite size and lattice parameter are found by reitveld fit and listed in table I. The discrepancy index for reitveld fit can be given by weighted profile R-factor ($R_{wp}$). The $R_{wp}$ is found as 2.16%, 2.64%, 2.59% and 2.44% for particle samples MF 1, MF 2, MF 3 and MF 4 respectively. The lattice parameters found are slightly lower than the typical value 0.839 nm for bulk magnetite [33]. Due to the large surface area, the $Fe^{2+}$ ions on surface can be oxidized to form the maghemite layer on the surface. This may be a possible reason for reduction of lattice parameter. But the presence of maghemite must be in very low proportion and the corresponding XRD peaks are not visible. Such reduction in lattice parameter is previously reported in [34]. The crystallite size is smallest for MF 1 and largest for MF 4. The crystallite size of magnetite is controlled in our experiment by controlling synthesis temperature and pH. The effect of these parameters on crystallite size of nanoparticle in coprecipitation method is discussed in detail in reports [28] [34].

Table I Magnetic fluid parameters determined using XRD, TEM and Magnetization measurements.

| Sample name | XRD | | TEM | | Magnetization | | |
|---|---|---|---|---|---|---|---|
| | Particle size ($D_{XRD}$) (nm) | Lattice parameter (nm) | Particle size ($D_{TEM}$) (nm) | Standard deviation $\sigma$ | Particle size ($D_{MAG}$) (nm) | Standard deviation $\sigma$ | Saturation Magnetization $M_S$ (Oe) |
| MF 1 | 10.60 | 0.8360567 | 11.86 | 0.21 | 10.5 | 0.26 | 141 |
| MF 2 | 12.0 | 0.83753 | 12.80 | 0.35 | 11.72 | 0.22 | 145 |
| MF 3 | 16.11 | 0.8373271 | 15.36 | 0.20 | 12.75 | 0.22 | 144 |
| MF 4 | 17.09 | 0.8352231 | 16.11 | 0.22 | 13.34 | 0.24 | 163 |

Fig.3 shows TEM images and particle size distribution for all four samples. The size distribution is fitted by lognormal distribution and, mean size and standard deviation are listed in table I. The mean size found from TEM analysis ($D_{TEM}$) is a physical or hydrodynamic size of particles. Fig.4 shows the magnetization measurement data fitted to modified Langevin's theory. The magnetization can be explained by Langevin's theory of paramagnetism (relation 1) for a monodispersed system. As magnetic fluid is a polydispersed system, Langevin's theory is modified to consider particle size distribution as described by relation 2. In modified theory, Langevin function $L\left(\frac{mH}{k_BT}\right)$ is weighted by lognormal size distribution function $F(D)$ given in relation 3. The modified Langevin theory is described in detail in [35].

$$\frac{M}{M_S} = L(\alpha) = \coth \alpha - \frac{1}{\alpha}; \ \alpha = \frac{mH}{k_BT}; m = M_d V \quad \ldots(1)$$

$$\frac{M}{M_S} = L(\alpha)F(D)dD \quad \ldots(2)$$

$$F(D)dD = \frac{1}{\sqrt{2\pi}\sigma D} exp\left[\frac{-(lnD - lnD_0)^2}{2\sigma^2}\right] dD \quad \ldots(3)$$

where $M_S$ is saturation magnetization of magnetic fluid, $m$ is particle magnetic moment, $H$ is magnetic field strength, $k_B$ is boltzman constant, $T$ is temeparature, $M_d$ is saturation magnetization of the bulk material, $V$ is particle volume, $D_0$ is mean particle diameter and $\sigma$ is standard deviation. The mean particle size, standard deviation and saturation magnetization of magnetic fluid can be found by best fitting of the experimental data to the modified theory.

The values obtained for fitting parameters saturation magnetization ($M_S$), mean diameter ($D_0 \approx D_{MAG}$) and standard deviation ($\sigma$) for size distribution of particles are listed in table I. The sizes $D_{MAG}$ are smaller than $D_{TEM}$ as it is size of magnetic core often called magnetic size of particles. The hydrodynamic size is always greater than the magnetic size of particles as it includes the thickness of coating layer. The saturation magnetization increases with particle size. Similar results are reported in [28] [34] [36].

The complex magnetic permeability ($\mu^*$) has two components, real ($\mu'$) and imaginary ($\mu''$) and is given by $\mu^* = \mu' - i\mu''$. In the equilibrium state, magnetic moments existing in magnetic fluid are all randomly oriented. When magnetic fluid is influenced by EM wave (radio- microwave), magnetic moments get polarized by the magnetic field component of EM wave. The $\mu'$ component is a contribution from the magnetization that is in phase with alernating magnetic field and it depends on the extent of magnetic polarization. While the $\mu''$ component is a contribution from the magnetization that is out of phase with alternating magnetic filed and is related to loss. The occurrence of relaxation and resonance is expected. There are two relaxation mechanisms Brownian and Neel's mechanism. The particle to which moment is embedded physically rotates in the former case while moment itself rotates inside the particle in the latter case. The relaxation time for both of the mechanisms and the effective relaxation time ($\tau_{eff}$) can be calculated as,

$$\tau_{eff} = \frac{\tau_B \tau_N}{\tau_B + \tau_N} \quad \ldots(4)$$

$$\tau_B = \frac{3V'\eta}{k_B T} \quad \ldots(5)$$

$$\tau_N = \tau_0 \exp(\sigma) \cdot \sigma^{-1/2} \quad if\ \sigma \geq 2$$
$$\phantom{\tau_N} = \tau_0 \sigma \quad if\ \sigma \ll 1 \quad \ldots(6)$$

where $\tau_{eff}$ is effective relaxation time, $\tau_B$ is Brownian relaxation time, $\tau_N$ is Neel relaxation time, $\eta$ is viscosity of carrier liquid, $V'$ is hydrodynamic volume of particle, $\tau_0$ is precessional damping time (≈10$^{-9}$sec), $\sigma = KV/k_B T$, $K$ is anisotropy constant and $V$ is magnetic volume. The effect of particle size on effective relaxation time is discussed in [37]. The $\tau_{eff}$ increases with particle size. At absorption frequency ($f_{max}$), $\mu''$ attains a maximum where $f_{max}$ corresponds to $\tau_{eff}$. The $\mu''$ peak signifies occurrence of energy loss (absorption), and often called as loss peak. In equilibrilium, a magnetic moment is oriented in the direction of anisotropy field ($H_A$).

The incidence of EM wave causes a small disturbance and magnetic moment starts to preccess around $H_A$. If external magnetic field ($H_{ext}$) is applied, it will be added to $H_A$. When the frequency of precessional motion matches with the frquency of EM wave, the precession would be continued by absorbing energy from EM wave. It is called ferromagnetic resonance which leads to strong energy absorption in the system. At the resonance frequency ($f_{res}$), $\mu' = 1$. The f$_{res}$ can be given by,

$$f_{res} = \frac{\gamma}{2\pi}(H_A + H_{ext}) \quad -(7)$$

where $\gamma$ is gyromagnetic ratio of electron, $H_A$ is anisotropy field given by $H_A = 4K/M_S$, $H_{ext}$ is external static magnetic field. The frequency dependence of complex magnetic permeability for all four fluids in the absence of any external field is shown in fig.5. As the frquency increases, field alters it's direction much faster and the dipoles remain unresponded. So the extent of magnetic polarization and real component ($\mu'$) decreases with frequency as observed in fig.5a. As the $\mu'$ component drops, the $\mu''$ component increases with frequency and attain maximum (fig.5b). The initial permeability can be given by,

$$\mu_{ini} = 1 + \chi_{ini} = 1 + nm^2/3k_B T\mu_0 \quad -(8)$$

where $\chi_{ini}$ is initial susceptibility, $m$ is magnetic moment; $m = M_S V$, $M_S$ is saturation magnetization, $n$ is particle number density, $\mu_0$ is vacuum permeability.

The initial susceptibility is proportional to particle volume. The $\mu_{ini}$ (@0.25 GHz) is expected to increase with particle size which can be observed in our results fig.5a.

The $f_{res}$ is observed to be 1.28 GHz, 1.45 GHz, 1.62 GHz and 1.99 GHz for MF 1, MF 2, MF 3 and MF 4 respectively (fig.5). The $f_{max}$ is observed to be 1.26 GHz, 1.30 GHz, 1.42 GHz and 1.45 GHz for MF 1, MF 2, MF 3 and MF 4 respectively (fig.5). Both the $f_{res}$ and $f_{max}$ increase as the particle size increases in the fluid. The $f_{res}$ is directly proportional to anisotropy field ($H_A$) when $H_{ext} = 0$ as in relation 7. The $f_{res}$ is increasing with particle size leads to the possibility that anisotropy field ($H_A$)) and so the anisotropy constant ($K$) are also incresing with size in the concerned size range. Some reports [38] [39] say that anisotropy for the nanoparticle is not purely of volumetric origin and dominated by surface contribution due to large surface to volume ratio and, thats why anisoptopy constant for nano materials are very often larger than that of bulk material. According to that anisotropy constant decreases with particle size for the nanoparticles. Our results do not follow this approach. Here the anisotropy constant, includes the effects from magnetocrystalline nature, size, shape and interparticle interaction [40] and is called effective anisotropy constant $K_{eff}$. The nanoparticle synthesized here are not perfectly spherical, but they are slightly elongated which can be observed in TEM images, so shape anisotropy contributes to constant $K_{eff}$ [41]. The fluid with large mean particle size must be having large magnetic interactions between particles. All these factors contribute to the constant $K_{eff}$. In previous report [42], researchers have suggested that oleic acid molecules covalently bonded to the particle surface effectively reduces the surface spin disorder and the anisotropy is dominated by volume contribution in oleic acid coated magnetite nanoparticles. The effective anisotropy constant ($K_{eff}$) increases with particle size most probably due to shape effects and reduced surface spin disorder. Our results support this idea proposed in [42]. The increase in constant $K_{eff}$ by increasing particle size is also reported in [20] for oleic acid coated magnetite nanoparticle. The enhencement in magnetic properties and in resonance effect occurs by increasing particle size. The resonance effect contribution can be responsible for the rise in loss peak and it's shifting toward higher frequencies. The theoratical study [24] says that, $f_{res}$ and $f_{max}$ should increase as $\sigma$ increases where $\sigma = KV/k_B T$. Either increasing constant K or particle volume $V$ can increase $\sigma$ and it will lead to increased $f_{res}$ and $f_{max}$. The interparticle interactions may contribute to the $f_{res}$ only. According to a previous report [43], the interparticle interactions increase the $f_{res}$ considerably while the $f_{max}$ remains unaffected. This can be one possible reason for the large increment in $f_{res}$ and small increment in $f_{max}$ for the same change in particle size.

The complex magnetic permeability is determined under the influence of static magnetic field with strength 0-368 Oe for all four fluids. Results are presented in fig.6 and 7. On the application of static magnetic field, magnetic moments try to align in the field direction and form chain structures. The magnetic fluid is a polydispersed system, having some smaller as well as larger particles compared to the mean size. At lower field strength, larger particles will be the first affected and will align in chain structures. As the field strength increases, more and more particles align in chains and few particles will be left to rotate freely. But the magnetic moments can still overcome the anisotropy energy barrier ($KV$) and relax via Neel's mechanism without the physical rotation of particles. As the field strength increases, the energy barrier $KV$ increases and less magnetic moments are able to cross the barrier and relax via Neel's mechanism. This will lead the extent of magnetic polarization and so the magnetic permeability to decrease with field strength. It can be observed that $\mu'$ and $\mu''$ decreases with field strength at the lower frequency end in fig.6 and 7 respectively (for $H_{ext}$>60 Oe).

The $f_{res}$ shifts to higher frequency as the field strength increases in each of the fluids (fig.6). It is expected according to the relation 7. The field profile for MF 4 appears to be much different from MF 1 (fig.6). The $f_{res}$ shifts to 2.03 GHz at 368 Oe of field strength from 1.28 Gz at absence of field in MF 1. While it shifts to 2.97 GHz at 368 Oe of field strength from 1.99 GHz at absence of field in MF 4. For the same increment of field strength (368 Oe), shifting of the $f_{res}$ is larger in MF 4 compared to that in MF 1. The larger $f_{res}$ spreading bandwidth in MF 4 signifies the larger value of $H_A$ and the corresponding constant $K_{eff}$. For the 60 Oe of applied field strength, there is a large rise in $\mu'$ curve for fluid MF 1 (fig.6). While in case of fluid MF 4, the rise in $\mu'$ curve is comparatively small for 60 Oe (fig.6). When particle is influenced by magnetic field strength comparable or greater than it's anisotropy field then only the magnetic moment rotates in the direction of external field and aligned to form chain structures. The large rise in the $\mu'$ curve in MF 1 is due to the alignment of moments. The report [44] suggests that size and shape distribution of nanoparticles leads to the wide distribution of anisotropy constant ($K_{eff}$) and of $H_A$ in the system. There must be more number of particles in MF 1 having $H_A$ comparable or less than 60 Oe. In opposite to that in MF 4 there are very less number of particles having constant $H_A$ comparable or less than 60 Oe.

The loss peak and $f_{max}$ shifts to higher frequency as the field strength increases in each of the fluids (fig.7). As the field strength increases, contribution from resonance effect increases and the contribution from relaxation decreases. The $f_{max}$ approches $f_{res}$ with increasing field strength. In case of MF 1 and MF 2, $\mu''$ peak amplitude increases with field strength (up to 368 Oe) (fig.7). While in MF 3 and MF 4, amplitude increases up to 190 Oe and 118 Oe respectively, after that it starts to decrease (fig.7). As the field strength increases, initially the loss peak amplitude increases with field strength because of the presence of aligned magnetic moments. This increment will continue upto a critical field strength. Beyond that as the field strength increases, barrier $KV$ increases and less number of particles can participate in relaxation. So the loss peak amplitude decreases with field strength. The critical field must be higher for MF 1 and MF 2 because of smaller mean particle size. Because the most fine particles present in fluid will contribute to Neel's relaxation.

The magnetic loss tangent ($\tan\delta$) can be given by, $\tan\delta = \mu''/\mu'$ where $\delta$ is a loss angle between two magnetic permeability components. The $\mu'$ and $\mu''$ components corresponds to the loss less and lossy responses of material respectively. The magnetic loss tangent is a ratio of lossy to the loss less response invovled in complex magnetic permeability. It represents the loss-rate of energy for a dissipating system when applied energy in form of alternating electromagnetic field [45]. The reflection loss ($RL$) is the measure of the energy reflected back, can be calculated as,

$$RL\ (dB) = 20\ log\left|\frac{Z_{in}-Z_0}{Z_{in}+Z_0}\right| \quad .....(9)$$

Detailed derivation for $RL$ is given in [46]. For good microwave absorption properties of a material, it is desirable to have high loss tangent and low reflection loss. The $\tan\delta$ and $RL$ are calculated for the absence of static magnetic field for all four fluids and plotted in Fig.8. The maximum $\tan\delta$ is attained at 1.44, 1.46, 1.46 and 1.47 GHz for MF 1, 2, 3 and 4 respectively. The maximum $\tan\delta$ increases with mean particle size and is largest for MF 4. The minimum $RL$ is attained at 1.83, 1.87, 1.87 and 1.94 GHz for MF 1, 2, 3 and 4 respectively. The minimum $RL$ decreases with mean particle size and the lowest for MF 4. The maximum loss tangent and minimum reflection loss is achieved in MF 4 due to the maximum complex magnetic permeability as a result of larger mean particle size. The reflection loss is tabulated in table II for frequencies 1, 2 and 3GHz for four fluids in the absence of external magnetic field. Reduction in RL due to the particle size increment is largest at 3 GHz.

Table II Reflection loss (RL) for four magnetic fluids in the absence of external magnetic field.

| Sample name | $RL(dB)$ | | |
|---|---|---|---|
| | f=1GHz | f=2GHz | f=3GHz |
| MF 1 | -1.17116 | -1.99704 | -0.91576 |
| MF 2 | -1.27275 | -2.35231 | -1.6774 |
| MF 3 | -1.19656 | -2.37967 | -1.6818 |
| MF 4 | -1.33089 | -2.80922 | -2.26055 |

The $\tan\delta$ and $RL$ are also calculated for the influence of static magnetic field of strength 0-915 Oe. The results for $\tan\delta$ and RL are shown in figure 9 and 10 respectively. The frequency and field dependence of $\tan\delta$ is similar to the $\mu''$ component. The maximum $\tan\delta$ increases with field strength up to a critical strength and then decreases with field strength. The minimum $RL$ decreases as the field strength increases, after a certain field strength, it seems that the minima is shifted to a higher frequency beyond our instrumental range. It can be observed that for MF 4, $RL$ <-3dB in the approximate range 2.2-3 GHz at field strength 510 Oe. According to the relationship between relection loss and absorbed energy suggested in [47], when $RL$ < -3dB, almost 50% of energy is absorbed in the system.

From the results it is clear that, this kind of fluid can be used as wide bandwidth absorber. At a particular frequency, $\tan\delta$ and $RL$ can be fine-tuned by controlling the field strength.

The $f_{res}$, $f_{max}$, maximum $\tan\delta$ is observed to increase by 55.6%, 15% and 25.2% respectively and minimum $RL$ is observed to decrease by 34.5 % by increasing the mean size of Fe$_3$O$_4$ nanoparticles from 11.8 nm to 16.1 nm in magnetic fluid.

IV. Conclusion

The Magnetic fluids having Fe$_3$O$_4$ nanoparticles of varying mean size between 11 to 16 nm have been synthesized using chemical co-precipitation method. The frequency dependant complex magnetic permeability is reported for these four Magnetic fluids in the frequency range 250 MHz to 3 GHz. The initial permeability and frequency dependent complex permeability increases by increasing particle size in the fluid. The ferrimagnetic resonance frequency ($f_{res}$), absorption frequency ($f_{max}$) and loss tangent (tan δ) increases while reflection loss ($RL$) decreases with increasing mean particle size in the fluid. Increasing particle size leads to interparticle interactions and anisotropy energy (KV) to increase which is responsible for these results. The field dependence of these properties have also been studied. By controlling the mean particle size and strength of static magnetic field, it is possible to fine tune the frequency of resonance and maximum absorption, reflection loss, absorption, and other dielectric properties of magnetic fluid which are usually desirable in radio-microwave devices and other applications like Hyperthermia.

ACKNOWLEDGEMENTS

MJ acknowleges Inspire Fellowship Programme (IF140928), Department of Science and technology (DST), New Delhi for financial support. Authors thank CSIR-Central Salt and Marine Chemicals Research Institute (CSMCRI), Bhavnagar for providing X-ray diffraction and TEM analysis facility.

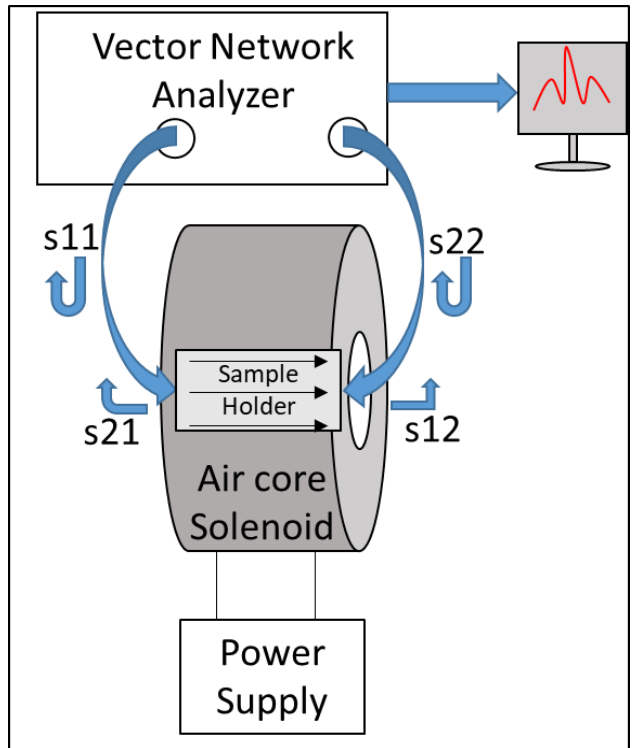

Figure. 1 Schematic of experimental set-up.

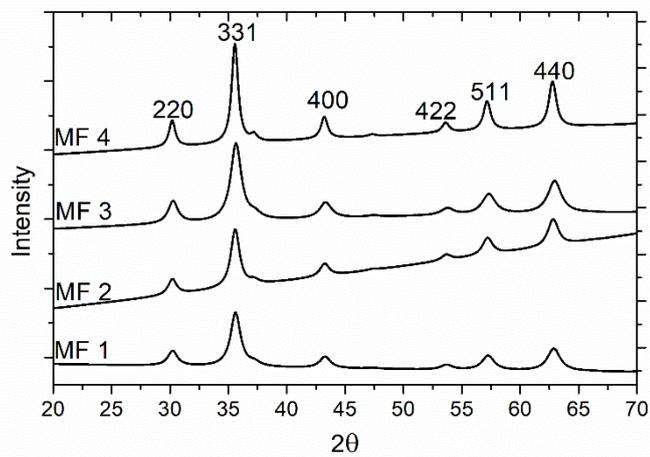

Figure.2 X-ray diffraction results for four samples of $Fe_3O_4$ powder.

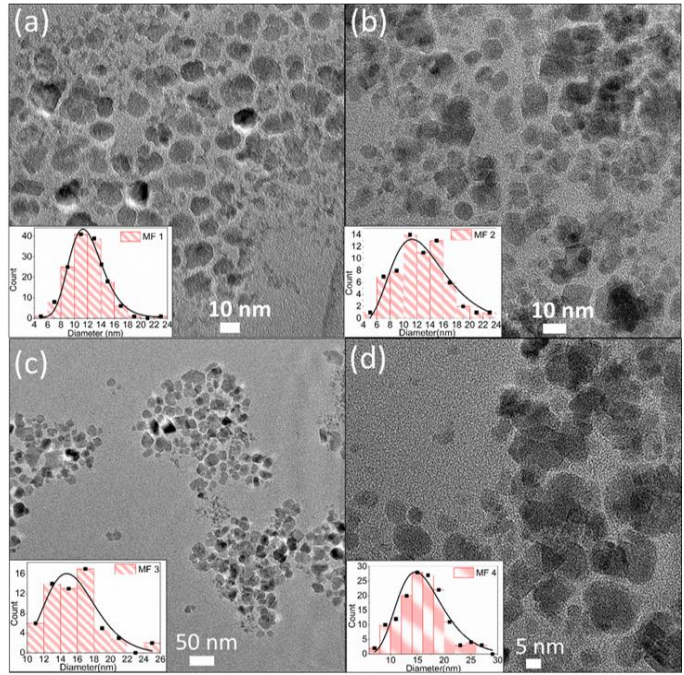

Figure.3 Transmission Electron microscope (TEM) images for four samples of $Fe_3O_4$ powder.

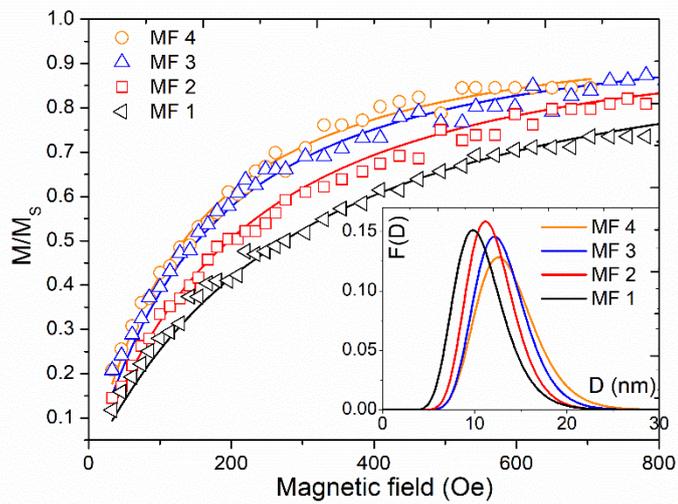

Figure.4 Magnetization measurement for four magnetic fluids. Hollow symbols represent experimental data points and lines represent best fit to modified Langevin theory. Inset shows lognormal particle size distribution in four fluids found from theory fit.

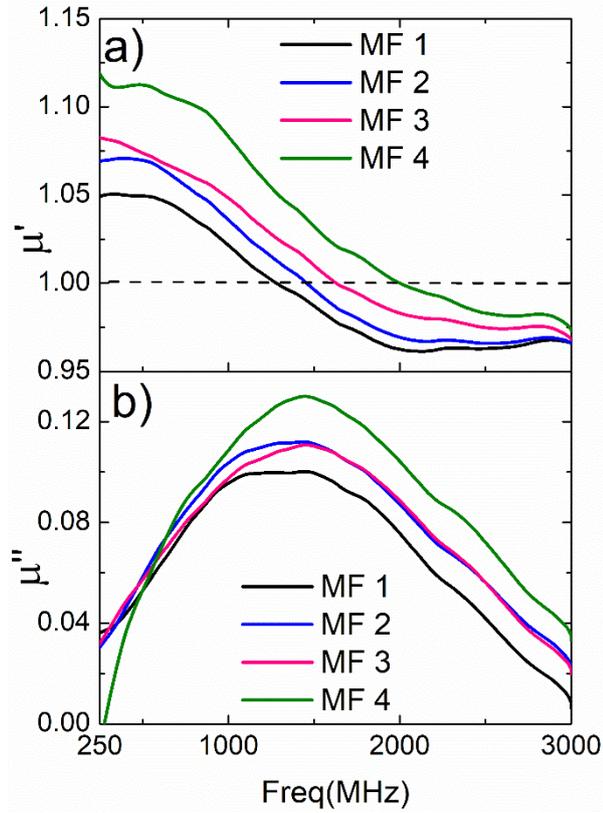

Figure.5 (a) The real and (b) imaginary components of complex magnetic permeability is plotted with frequency for four magnetic fluids in the absence of static field.

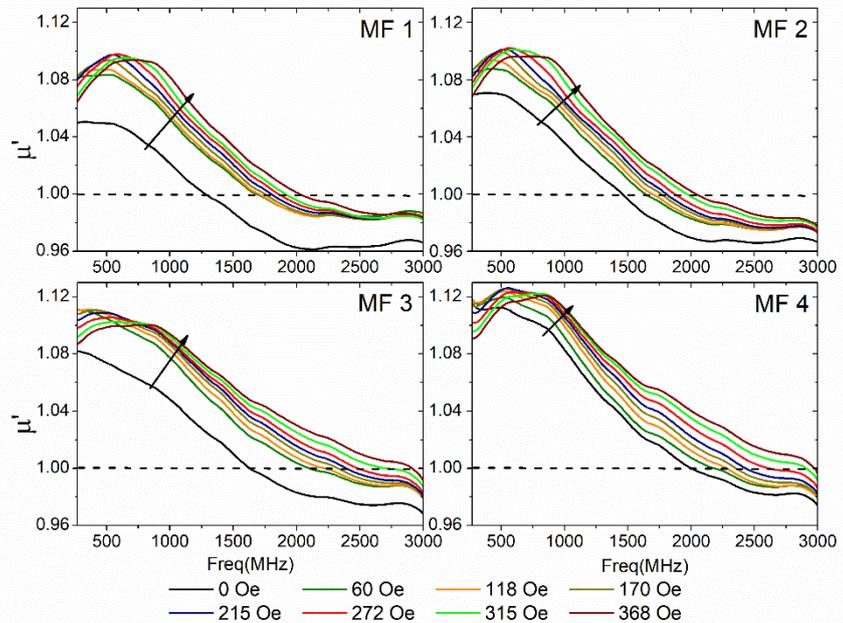

Figure.6 The frequency dependence of real component of complex magnetic permeability of MF 1, MF 2, MF 3 and MF 4 in the presence of static magnetic field with strength between 0-368 Oe.

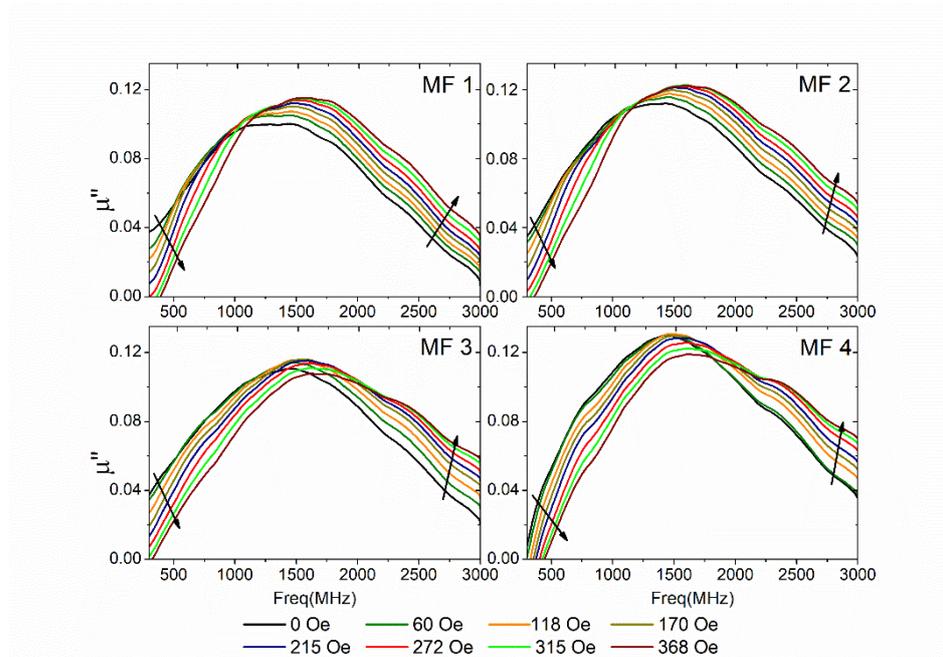

Figure.7 The frequency dependence of imaginary component of complex magnetic permeability of MF 1, MF 2, MF 3 and MF 4 fluids in the presence of static magnetic field with strength between 0-368 Oe.

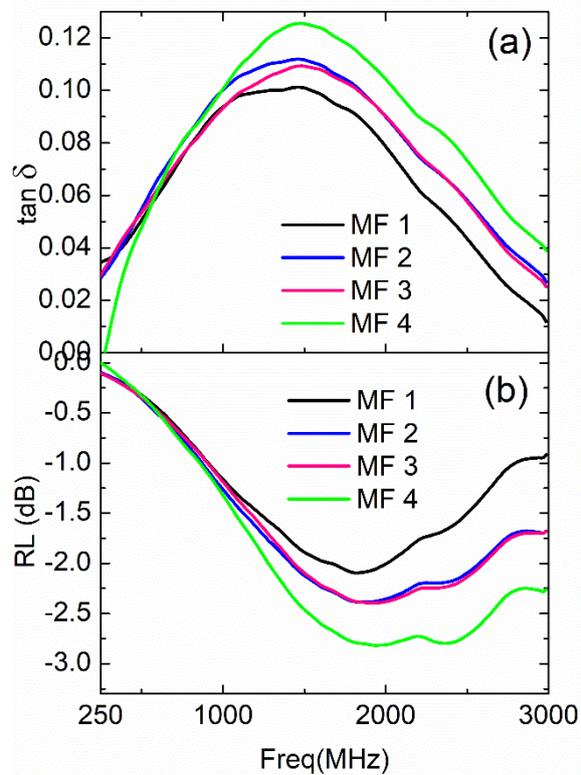

Figure.8 (a) Loss tangent and (b) Reflection loss is plotted with frequency for four fluids in the absence of static field.

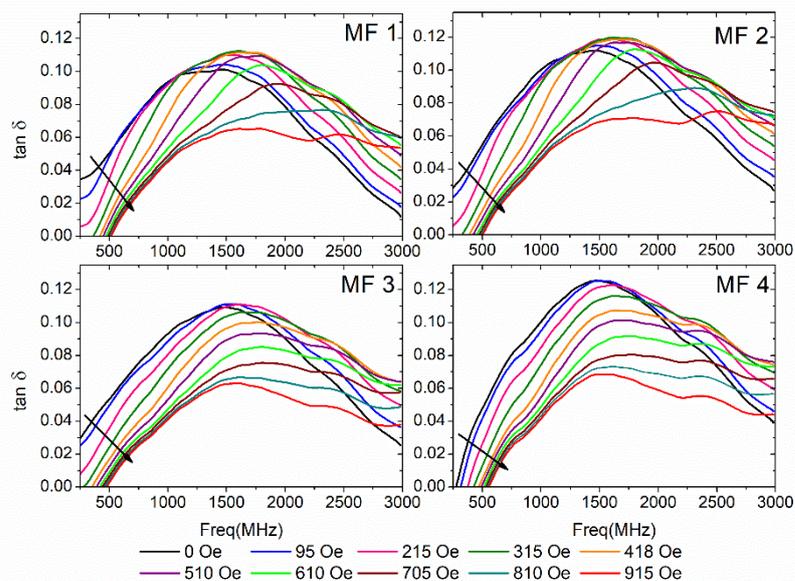

Figure.9 Frequency dependence of loss tangent (tan δ) for MF 1, MF 2, MF 3 and MF 4 fluids in the presence of static magnetic field with strength between 0-915 Oe.

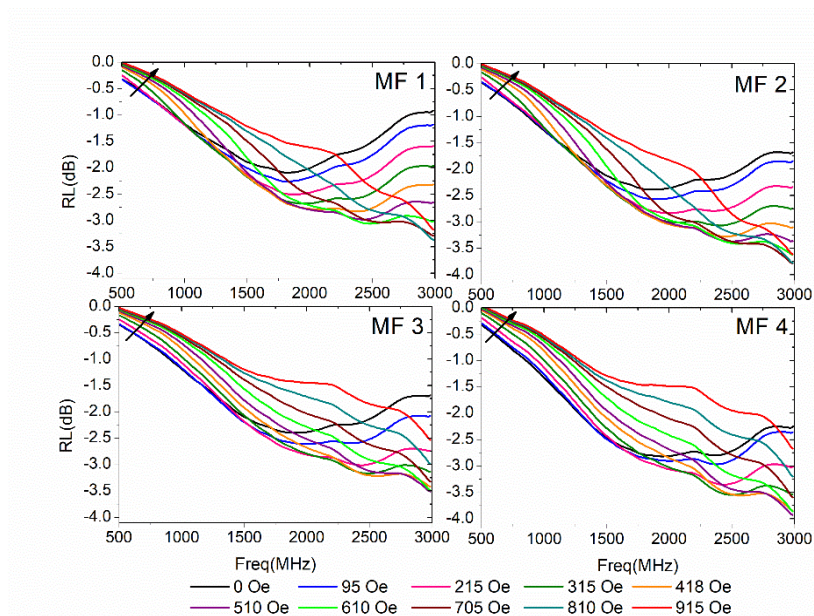

Figure.10 Frequency dependence of reflection loss (RL) for MF 1, MF 2, MF 3 and MF 4 fluids in the presence of static magnetic field with strength between 0-915 Oe.